# Magnetically Enhanced Fenton-Like Processes by Nanofibers: Real-Time Observation of Tetracycline Degradation in Pig Manure Wastewater.

**Berta Centro Elia** [1], **Marco Antonio Morales** [1,2], **Vanina G. Franco** [1], **Jesús Antonio Fuentes García**[1,3] **and Gerardo F. Goya** [1,3,*]

[1] Instituto de Nanociencia y Materiales de Aragón (INMA), CSIC-Universidad de Zaragoza,
Campus Río Ebro, 50018 Zaragoza, Spain.
[2] Div. Resonancias Magnéticas, Centro Atómico de Bariloche/CONICET, S.C 8400, Bariloche, Argentina.
[3] Departamento de Física de la Materia Condensada, Facultad de Ciencias, Universidad de Zaragoza,
50009 Zaragoza, Spain
* Correspondence: goya@unizar.es;

**Abstract:** This study presents a novel approach for the degradation of tetracycline (TC) in pig manure wastewater using magnetite-based magnetic nanofibers (MNFs) as heterogeneous Fenton-like catalysts. The MNFs, composed of polyacrylonitrile (PAN) embedded with $MnFe_2O_4$ nanoparticles, were synthesized via electrospinning and exhibited high stability and catalytic efficiency. The degradation process was driven by hydroxyl radical (•OH) formation through hydrogen peroxide ($H_2O_2$) activation on the MNF surface. The results showed that TC was first adsorbed onto the MNFs before undergoing oxidation, with treatment efficiency increasing with $H_2O_2$ concentration up to an optimum point, due to increased •OH scavenging by $H_2O_2$. A heterogeneous dynamic kinetic model (DKM) was developed to describe the degradation mechanism, incorporating reactive oxygen species (ROS) generation, catalyst surface inactivation, and polymer stripping effects. Furthermore, the application of an alternating magnetic field significantly accelerated the reaction rate, likely due to localized heating effects. This study highlights the potential of MNFs as a scalable, reusable and efficient alternative for antibiotic-contaminated wastewater treatment, offering advantages over conventional homogeneous Fenton processes by minimizing iron sludge formation and broadening the operational pH range.

# 1. Introduction

Water pollution, driven by industrial development and population growth, is a major environmental concern, and consequently a priority under the UN's Sustainable Development Goals (SDGs).[1] The problem is exacerbated by the discharge of wastewater contaminated with various products from urban and agricultural activities (hospitals, pharma, domestic, livestock farming) into water bodies.[2] These pharmaceuticals belong to the group of so-called emerging contaminants (ECs), with concentrations ranging from ng/L to mg/L. Tetracycline (TC), a widely used antibiotic, is prevalent in both human medicine and livestock feed,[3] and as a consequence of this pervasive use TC-contaminated wastewater poses a significant ecological risk and a threat to human health.[4] Consequently, addressing aqueous antibiotic pollution, particularly TC pollution, has become an urgent public concern.

Among various wastewater treatment technologies, advanced oxidation processes (AOPs)[5] have garnered significant attention for their ability to eliminate refractory pollutants by generating highly reactive species such as •OH (hydroxyl radical).[6] The Fenton process operates as an AOP under normal temperatures and pressures and is initiated by mixing $Fe^{2+}$ with $H_2O_2$ to produce •OH, with $Fe^{3+}$ also catalyzing $H_2O_2$ to regenerate $Fe^{2+}$, through the following reactions [7]

$$Fe^{2+} + H_2O_2 \rightarrow Fe^{3+} + 2 \cdot OH \ (K_1 = 70 \ M^{-1} s^{-1})$$

$$Fe^{3+} + H_2O_2 \rightarrow Fe^{2+} + H^+ + \cdot HO_2 \ (K_1 = 0.001 - 0.1 \ M^{-1} s^{-1})$$

Despite the advantages, traditional homogeneous Fenton reactions have limitations, such as a strict pH requirement around 3, high costs for effluent neutralization, and iron sludge generation. To address these drawbacks, researchers have developed heterogeneous Fenton-like processes using solid iron-based catalysts instead of soluble iron salts.[8]

The livestock industry, particularly the pork sector, is a major contributor to antibiotic contamination, with tetracycline (TC) being widely used. TC is a broad-spectrum antibiotic that inhibits bacterial protein synthesis and is highly persistent in soil and water due to its chemical stability. Its tetracyclic structure includes hydroxyl, ketone, and amine functional groups, contributing to its environmental resilience and resistance to degradation. [9] The Spanish pork industry, representing 39% of total livestock production, faces challenges in managing waste such as slurry, which contains antibiotic residues. Despite its environmental impact, there is no specific European regulation for TC in water. The EU's Water Framework Directive (WFD) monitors Priority Substances (PS), setting Environmental Quality Standards (EQS) to protect ecosystems. Given its structural similarity to regulated contaminants like PAHs, benzene, and benzo(a)pyrene, existing parametric values could serve as provisional references for TC. Although this criterion must be considered as a rule of thumb (reflecting the need for further research on TC environmental impact), Table 1 presents the maximum permissible concentrations for these related contaminants, which define legal limits for drinking water supplies.

Table 1

| Compound | Parametric value |
|---|---|
| Benzene | 10 μg/L |
| Benzo(a)pyrene | 0.010 μg/L |
| Polycyclic Aromatic Hydrocarbons (PAHs)* | 0.10 μg/L |

* Sum of concentrations of specified compounds: benzo(b)fluoranthene, benzo(k)fluoranthene, benzo(ghi)perylene, indeno(1,2,3-cd)pyrene

Magnetically activated degradation is a promising strategy for enhancing advanced oxidation processes (AOPs) in wastewater treatment. [10] Magnetic nanoparticles (MNPs), particularly those based on iron oxides such as magnetite ($Fe_3O_4$) have been extensively studied as heterogeneous Fenton-like catalysts due to their ability to generate reactive oxygen species (ROS) under ambient conditions.[11] [12] The application of an alternating magnetic field (AMF) induces localized heating through Néel relaxation mechanisms, accelerating reaction kinetics and enhancing the efficiency of pollutant degradation. Recent studies have demonstrated that magnetic hyperthermia can improve catalytic performance by increasing the availability of active sites and reducing mass transfer limitations, particularly in solid-phase catalysts.[13] Magnetically responsive materials, such as nanofibers embedded with MNPs, offer advantages in reusability and separation, minimizing secondary contamination. While traditional Fenton processes require acidic conditions and produce iron sludge, magnetically assisted catalysis operates over a broader pH range, making it a more sustainable alternative for treating complex wastewater matrices, including antibiotic-contaminated livestock effluents.[14]

In this work, we present an innovative system for treating pig manure-contaminated wastewater containing antibiotics, specifically TC, using magnetic polymeric nanofibers to degrade this organic pollutant through Fenton-like heterogeneous catalysis. The nanofibers, composed of polyacrylonitrile (PAN) and embedded with $MnFe_2O_4$, provided both the thermo-responsive action and the catalytic activity, accelerating the spontaneous kinetics of TC degradation when an external magnetic field was applied.

## 2. Materials and Methods

*Chemicals*

All chemicals used in this work were applied as purchased, without further purification. Fe(II) sulphate heptahydrate ($FeSO_4 \cdot 7H_2O$, ReagentPlus® ≥ 99%), Mn(II) sulphate monohydrate ($MnSO_4 \cdot H_2O$, ReagentPlus® ≥ 99%), sodium hydroxide pellets (NaOH, ACS reagent ≥ 97 %). Materials for the nanofibers included polyacrylonitrile (PAN, $(C_3H_3N)_n$) average Mw 150,000, N,N-dimethylformamide (DMF, $CHON(CH_3)_2$) (ReagentPlus® ≥99%). Milli-Q water (18.2 MΩ·cm) was used during sample preparation. Hydrochloric acid (HCl 37%, Labbox, Spain), nitric acid ($HNO_3$, 65%, Panreac), ammonium thiocyanate ($NH_4SCN$, Merck), Hydrogen peroxide ($H_2O_2$, Labkem, 30% v/v) and pure tetracycline ($C_{22}H_{24}N_2O_8 \cdot HCl$ ≥95%, Sigma Aldrich) were employed for the corresponding experiments.

*Synthesis of magnetic Nanoparticles (MNP)*

$MnFe_2O_4$ magnetic nanoparticles (MNPs) were synthesized by sonochemistry in aqueous solution as reported elsewhere.[15] Briefly, a solution of $Fe^{2+}$ and $Mn^{2+}$ (5.4 mM) was prepared in Milli-Q water in a 9:1 $FeSO_4$ ratio, placing 90 mL of the prepared solution in the reactor with the sonotrode. The solution was continuously ultrasonicated (0.27 J/mL) for 10 minutes, after which 10 mL of NaOH solution (2 N) was added at a rate of 20 mL/min. The reaction was maintained for an additional 10 minutes under continuous irradiation. The obtained MNPs were mixed and washed with Milli-Q water using magnetic decantation until the pH reached 7. The powdered MNP samples were obtained by drying in an air atmosphere at 60 °C for three days.

*Processing of magnetic Nanofibers (MNF)*

For fabrication of magnetic Nanofibers, 1 g of the prepared MNPs was re-dispersed in 10 mL of DMF using an ultrasonic bath for 10 minutes. The obtained dispersion was placed on magnetic stirring (250 RPM) and heated at 80 °C to improve the solubility of the polymer. Then, 1 g of PAN was dissolved in the DMF-MNP dispersion for 2 hours.[12] Heating was turned-off and stirring continued for 12 hours to homogeneous dispersion of MNP in the PAN matrix and leading to adequate entanglement of polymeric chains. Polymeric nets composed by PAN nanofibers and magnetic nanoparticles (labelled as MNFs samples hereafter) were processed injecting 1 mL of the previously prepared PAN-MNP solution within the electrospinning system. The feed rate was 0.7 mL/h, while distance needle-collector and high-voltage applied were 15 cm and 8 kV respectively (see Figure 1). The electrospun mats deposited on Al foil. Once the fibers were made, they were left to dry for a minimum of 12 hours so that the chains would compact, first at 60ºC in the oven (to dry without evaporating and affecting the structure) for about 3 hours and then they were left at 120-150ºC for 9 h . After the MNFs were dry, samples were conditioned by trimming the patches in 10 mm diameter circles for further characterization and evaluation of their hyperthermia response and degradation capacity.

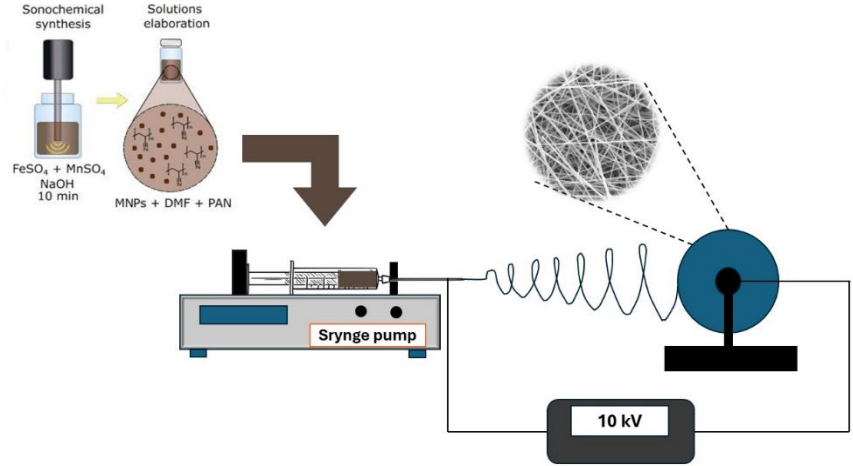

Figure 1. Schematic illustration of the synthesis and fabrication process of magnetic nanofibers. The sonochemical synthesis of $MnFe_2O_4$ magnetic nanoparticles is followed by the preparation of a polymeric solution (MNPs dispersed in DMF + PAN). This solution is then subjected to electrospinning using a syringe pump, applying a high voltage (10 kV), resulting in the formation of a nanofiber mat.

*Characterization*

The physical and chemical properties of the obtained MNPs and MNFs were characterized using various techniques. Morphology and particle size were analysed through Transmission Electron Microscopy (TEM) imaging with a FEG TECNAI T20 operating at 200 kV, and ImageJ software (version 1.56t) was used for image analysis. Diluted MNP dispersions were prepared, and a 10 µL drop was deposited onto Holley-carbon copper grids for observation and imaging. Additionally, imaging and energy dispersive spectroscopy (EDS) analysis was performed for MNFs with a Quanta FEG 650 Scanning Electron Microscope (SEM) equipped with a ThermoFisher® X-ray photon detector. Samples of the prepared electrospun fibers were coated with carbon for improved electron conductivity during the SEM observations.

*AC magnetic field measurements*

The Specific Loss Power (SLP) values of the MNPs were measured with a magnetic field applicator within the frequency range $180\ kHz\ \leq\ f\ \leq 765\ kHz$, in magnetic field amplitudes up to $H_0 = 50\ kA/m$. The heating capacity of the MNFs was determined using the calorimetric formula to ascertain the Specific Power Loss (SLP) as the primary physical parameter representing the heating capacity at a particular amplitude and frequency of the magnetic field. It is given by:

$$SLP = \frac{m_{H2O} \cdot c_{H2O} + m_{MNF} \cdot c_{MNF}}{m_{MNF}} \left(\frac{\Delta T}{\Delta t}\right)_{max}$$

where $m_{H2O}$ and $m_{MNF}$ are the masses of the carrier liquid (water) and the nanofibers, respectively, while $c_{H2O}$ and $c_{MNF}$ are the specific heat capacities of the liquid and the nanofibers (water, $c_{H2O} = 4186\ \frac{J}{kg \cdot K}$; and PAN $c_{MNF} = 1350\ \frac{J}{kg \cdot K}$).

A second Power Loss experiments were performed under ac magnetic fields ($H$ = 32 kA/m; $f$ = 350 kHz, Ambrell Ltd.) with the temperature evolution of the sample monitored through a commercial thermographic camera (resolution, 240 × 180 pixels; temperature accuracy, 2 K), and obtaining images and the temperature value of the surface of the MNFs patches during heating in air (i.e., without using water as a medium). For this, we recorded video sequences during heating and subsequently extracted the temperature values from the frames corresponding to different time intervals.

Preparation of Pig Farm Manure Samples

Approximately 1 liter of liquid waste was collected from a pig farm (Cuarte S.L., Grupo Jorge) 48 hours after antibiotic administration. The samples were transported in non-sterile containers and stored at 4 °C until processing. All laboratory materials were sterilized before use, and appropriate personal protective equipment (PPE) was worn throughout the process. To prepare the sample for laboratory analysis, large solids and suspended particles were first removed through centrifugation in 50 mL Falcon tubes at 10,000 rpm for 20 minutes at 4 °C. This process was repeated three times until a two-phase separation with a slight color difference was observed (Figure S2 in the Supplementary Material). The supernatant was carefully decanted into a new container, avoiding disturbance of the solid pellet, which was discarded. The supernatant was then filtered using a vacuum filtration system equipped with 0.45 µm membrane filters (PRAT DUMAS, model A007607) to remove fine particles and unwanted organic solids. The filtrate was collected in a clean container for further processing. To remove salts and other small impurities while concentrating the TC, a dialysis process was performed using dialysis tubes with a molecular weight cut-off (MWCO) of 100 kDa. These tubes were hydrated in distilled water for 30 minutes before use, then filled with the filtrate and placed in a container with Milli-Q water. Dialysis was carried out under gentle stirring at room temperature for 72 hours without changing the dialysis solution to allow TC accumulation until equilibrium was reached (Figure S3 of the Supplementary Material). Upon completion, the solution was removed for analysis, and the contents of the dialysis tubes were discarded.

*UV-vis determination of Tetracycline Degradation by MNFs*

In all the degradation experiments conducted, 4 disks of MNFs were pressed between two nylon membrane filters with a thickness of 0.45 µm and a diameter of 13 mm (Whatman). These filters act as mechanical support for the nanofibers. In all the degradation experiments, 2% hydrogen peroxide was added to the TC solution. The control experiments for TC degradation were conducted by preparing a standard solution of TC hydrochloride ($C_{22}H_{24}N_2O_8 \cdot HCl$) at a concentration of 1 mg/L. This solution was made in 50 mL of distilled water and stored in a covered bottle to prevent exposure to light. A UV-Vis calibration curve was performed to quantitatively convert absorption into TC concentration. The experimental setup including the light source and spectrophotometric detector, and the magnetic field applicator, is shown in Figure S4 of the Supplementary Material.

Tetracycline was determined using a UHPLC instrument (Waters Acquity) equipped with a PDA UV detector and a QDa mass detector (Waters Acquity). Samples were eluted on a Waters Cortecs T3 column (2.1 x 75 mm, 1.6 µm) at 40°C and a flow rate of 0.5 mL/min, using a gradient of a mixture of acetonitrile (ACN) and water, both doped with formic acid (0.1% v/v) as the eluent. The column gradient used is shown in Table 2.

Table 2. Acetonitrile/water gradients used for column elution to determine tetracycline.

| t (min) | ACN (% v/v) | H2O (% v/v) |
|---------|-------------|-------------|
| 0       | 5           | 95          |

| 1 | 5 | 95 |
|---|---|---|
| 14 | 95 | 5 |
| 15 | 95 | 5 |

Tetracycline was detected at 350 nm by UV, and in ESI$^+$-MS the ion at m/z = 445 [M+H]$^+$ was monitored.

## 3. Results

*3.1. Physicochemical characterization of MNFs*

The TEM analysis of the MNPs showed that the nanoparticles predominantly exhibit a cubic morphology, suggesting controlled and reproducible synthesis. Additionally, it can be observed that the nanoparticles are grouped relatively homogeneously, indicating good dispersion and stability within their medium. The size distribution analysis represented in the histogram (Figure 2) revealed an average diameter of 72 ± 14 nm, a good compromise between the optimal large size needed for power absorption and the smaller size that could be more efficient for larger specific surface area and thus more catalytic efficiency.

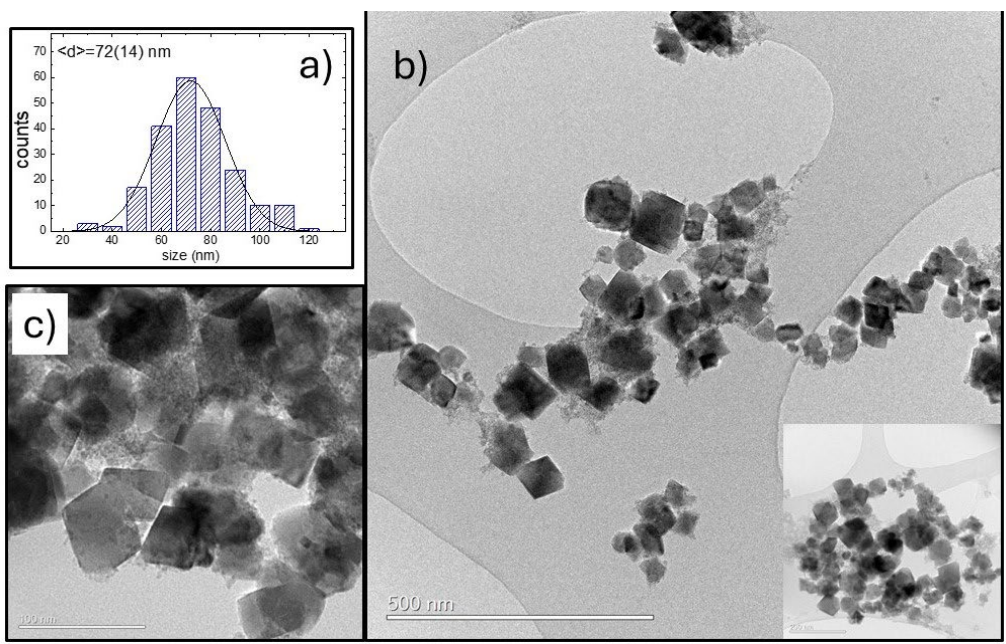

Figure 2. TEM images of the MnFe$_2$O$_4$ MNPs embedded into the MNFs. Panel a) Size distribution histogram fitted with a gaussian function of <d> = 72 ± 14 nm. Panels b) and c) show different magnification values.

Regarding the *as prepared* patches of MNFs (see Figure S1 in the Supplementary Material), SEM images revealed a homogeneous distribution of the MNPs within the polymeric material, together with the presence of some agglomerates on the fiber's surface (see Figure 3).

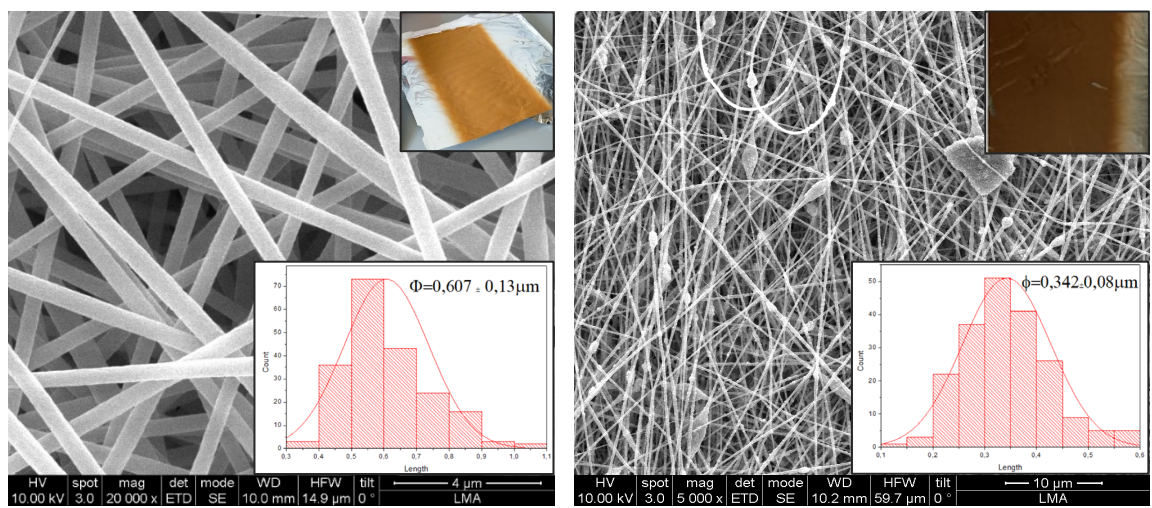

Figure 3. SEM images of PAN MNFs (left panel) and MNFs with MnFe$_2$O$_4$ nanoparticles (right panel).

Pure PAN-MNFs exhibited a uniform morphology with an average fiber diameter of 607 ± 13 nm, while MNP-containing MNFs had a rougher surface due to the partial distribution of MnFe$_2$O$_4$ MNPs. The fibers with nanoparticles had an average diameter of 342 ± 80 nm, indicating a reduction in size. SEM images (Figure 3) confirmed these diameters and showed that MNPs were both embedded within the fibers and distributed on the surface, occasionally forming agglomerates coated with a thin PAN layer. Energy-dispersive X-ray spectroscopy (SEM-EDS) confirmed the Mn:Fe ratio of 1:2, consistent with the expected MnFe$_2$O$_4$ stoichiometry (see **Table 3**) and verified the presence of manganese (Mn) and iron (Fe) within experimental error.

Table 3. Elemental composition of MnFe$_2$O$_4$ particles by EDS, at three sample areas.

|  | Sample 1 (at%) | Sample 2 (at%) | Sample 3 (at%) |
|---|---|---|---|
| Mn | 1,89 | 1,97 | 1,88 |
| Fe | 4,18 | 3,54 | 3,98 |
| Mn/Fe | 2.20 | 1.81 | 2.11 |
| stoichiometry | Mn$_{0.94}$Fe$_{2.06}$O$_4$ | Mn$_{1.13}$Fe$_{1.87}$O$_4$ | Mn$_{1.01}$Fe$_{1.99}$O$_4$ |

The heating capacity of the MNPs was assessed through Specific Loss Power (SLP), under a fixed magnetic field amplitude $H_0 = 32\ kA/m$ and frequency ($f = 350\ kHz$). The heating rate of the sample determined from the point of maximum slope of the T vs t curves (Figure S5 in the Supplementary Material) gave a value of $SLP = 325.5\ \frac{W}{g}$. Similar calculations with the MNFs (see Figure S7 in Supplementary Material) in air using the thermographic data yielded $\left(\frac{dT}{dt}\right)_{max} = 1.71\ \frac{K}{s}$ and a corresponding $SPL = 2295\ W/g$, demonstrating the excellent capacity of the MNFs material to respond to inductive heating. Since the MNPs are immobilized within the fibres, Brownian relaxation is blocked and therefore only Néel relaxation contributed to the SLP. In terms of temperature increase, MNFs in water showed a temperature increase of ΔT ≈ 20 °C after 5 min, while the direct measurement of the nanofibers in air yielded a notable ΔT ≈ 27 °C in only 30 s (see Figure S6 in

Supplementary Material), demonstrating the capacity of the MNFs to thermally activate the kinetics by local heating, as discussed above.

*3.2. Degradation of TC by MNFs*

To assess the effect of the addition of $H_2O_2$ on the UV-Vis absorption peaks of the TC compound, absorbance measurements were performed on TC solutions treated with $H_2O_2$ at different concentrations (12 M to 0.19 M). The spectrum of the untreated TC presents two characteristic peaks at $\lambda$ = 276 nm and $\lambda$ = 358 nm (Figure 4). By increasing pH values the 276 nm peak increases its intensity shifting to $\lambda \approx$ 250 nm, as it originates in different TC functional groups that are deprotonated with higher pH values. On the other side, the peak at $\lambda$ = 358 nm shows stability in all pH range, and the derivatives at this maximum $\lambda$ = 358 nm (insert of Figure 4) show that the position of the maximum changes by less than ≈ 1 nm over the entire pH range.

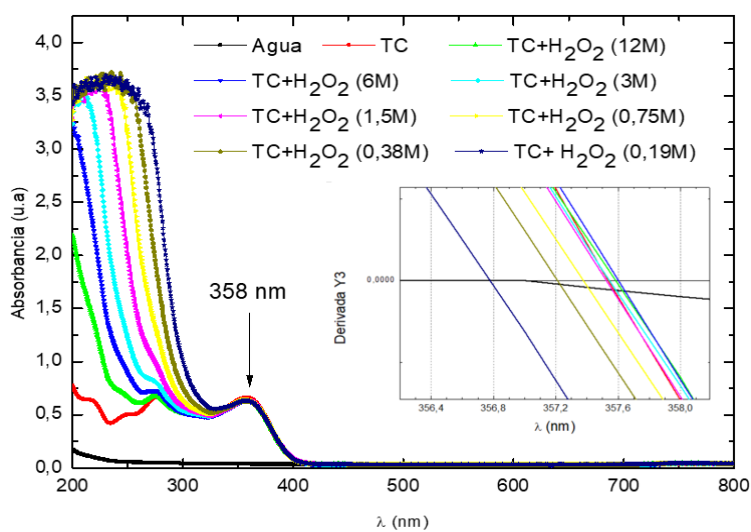

Figure 4. TC absorption spectra at different $H_2O_2$ concentrations. The insert shows the absorbance derivatives in the region of the maximum at $\lambda$=358 nm, demonstrating the stability of this peak with the pH change within a $\Delta\lambda \approx$1 nm shift.

From the above results, the measurement of TC concentrations during degradation was determined using the pH-independent peak at $\lambda$ = 358 nm. Subsequently, the degradation kinetics of pure TC by the MNFs was tested, and a typical dynamic profile obtained over time (using a ≈ 6 µg/mL TC solution) is shown in Figure 5. The effect of a "blank" polyacrylonitrile (PAN) fibers patch (i.e., without embedded nanoparticles) on TC with $H_2O_2$ was also measured as a blank control (blue triangles), finding a small decrease after 50 h in an approximately linear trend, at a rate of $\approx 8 \frac{ng}{mL \cdot hr}$. This can be attributed to spontaneous decomposition of TC under the action of the UV-Vis light in the spectrophotometer, to a partial adsorption by the polymer, or a combined effect. Nevertheless, this small effect observed for blank MNFs was considered negligible compared to the degradation decrease. Under identical conditions of [$H_2O_2$] and temperature, the degradation profiles of TC using MNFs (red curve), exhibited a three-phase kinetic behaviour. Initially ($t_{peel} \approx$ 5 h), the degradation rate remained relatively low, indicating a slow-kinetics phase where minimal tetracycline degradation occurs. We refer to this interval as "induction period". This is followed by a rapid decline in TC concentration within the following ≈ 20 h with enhanced catalytic activity likely driven by the active generation of reactive oxygen species (ROS). After this accelerated degradation phase, the reaction rate progressively decreases, transitioning into a slower reduction stage that stabilized near complete degradation of TC after approximately 40 hours.

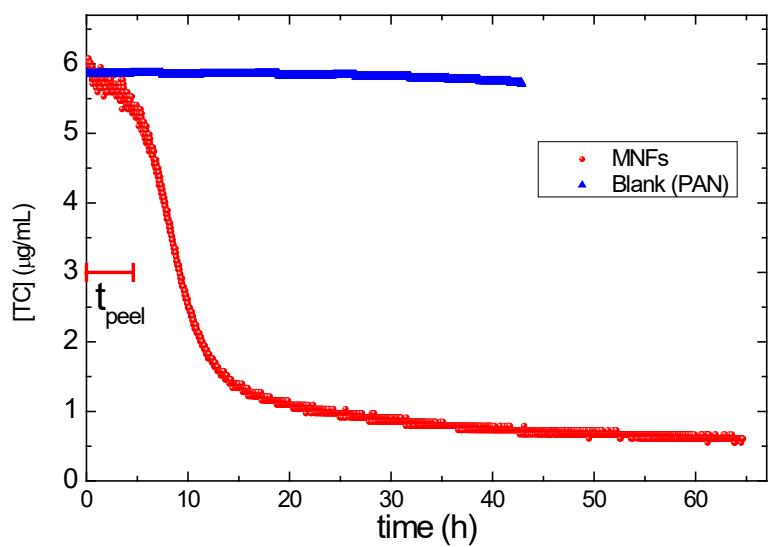

Figure 5. Evolution of TC degradation over time: blank (PAN) fibers used as control (blue triangles) showed a small constant degradation rate of $\approx 8 \frac{ng}{mL \cdot hr}$ along 50 h of the experiment, indicating negligible catalytic activity. In comparison, the MNFs (red circles) exhibited effective TC degradation, stabilizing near [TC] ≈≈50 ng/mL after approximately 40 hours.

This overall pattern suggests that MNFs serve as effective catalysts in the degradation of TC, most likely through a Fenton-like mechanism through the decomposition of $H_2O_2$, leading to the generation of highly reactive •OH, which are primarily responsible for breaking down TC molecules in solution. The oxidative degradation observed therefore occurs with a distinct interplay between slow initiation, rapid decomposition, and eventual stabilization as reactive species become depleted or surface catalytic sites are passivated.

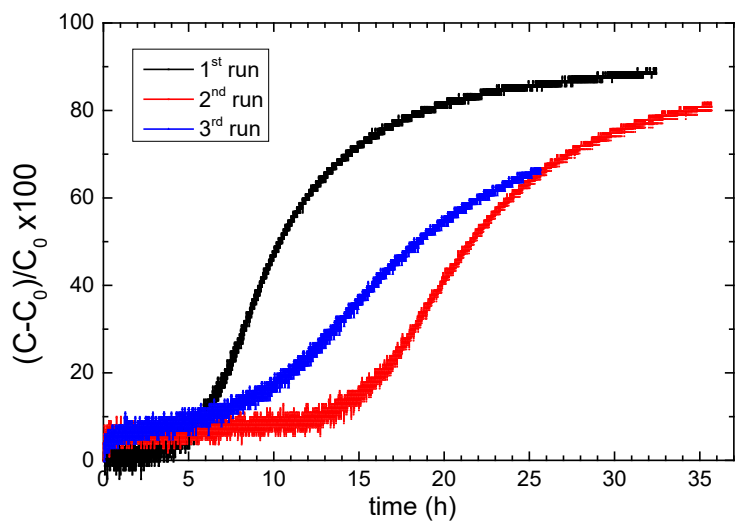

Figure 6. Degradation efficiency of the MNFs over three consecutive reuse cycles. The first run (black curve) achieved approximately **90%** degradation, while the second (red) and third (blue) runs showed a gradual decrease in efficiency, reaching around **67%** in the third cycle. This decline suggests partial deactivation of catalytic sites or material loss, yet the MNFs retain a significant degradation capability, demonstrating their potential for reuse in catalytic applications.

The MNFs demonstrated a notable capacity for reuse over multiple cycles while maintaining an acceptable degradation performance. As observed in Figure 6, the degradation efficiency in the first run reached approximately 90%, indicating a highly effective catalytic process. However, with subsequent reuse, a gradual decrease in degradation efficiency was observed, dropping to around 67% after three cycles. This decline suggests a possible reduction in catalytic activity, likely due to factors such as partial deactivation of active sites, material loss, or structural modifications over time. Despite this decrease, the MNFs still exhibited a significant degradation capability, highlighting their potential for multiple uses in catalytic applications with reasonable efficiency retention.

*3.3. Effect of magnetic field-assisted degradation*

The comparative data from the degradation experiments with and without an applied magnetic field are shown in Figure 7. The comparison between the spontaneous degradation of TC by the MNFs and the degradation under an alternating magnetic field (H = 32 kA/m; $f$ = 450 kHz, applied for two successive intervals of one hour each), revealed markedly different behaviors. In the absence of the field, an initial 'inactive' period of approximately 4–6 hours was observed. In contrast, upon application of the magnetic field, the degradation process began within ~1 minute after field activation ($H_{ON}$ in the inset of Figure 7), significantly faster than the ~5-hour induction period observed without the field. Furthermore, the degradation rate following $H_{ON}$ was noticeably speed up. This faster reaction rate could be due to enhanced catalytic activity induced by the field on the active sites and/or the effect from the heating during the field applications. As time progresses, the degradation under the magnetic field slows down, approaching a plateau after t ≈ 12 h, as most of the TC has been degraded.

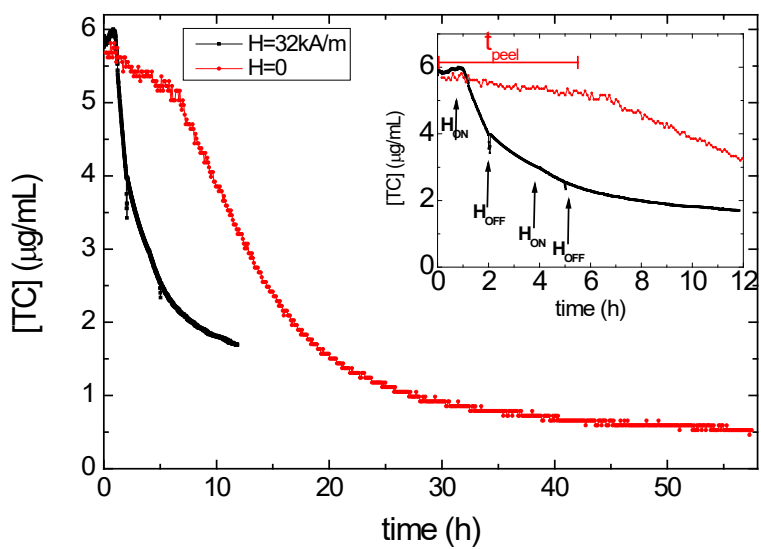

Figure 7. TC degradation profiles by MNFs: a) spontaneous degradation without applied magnetic field (red circles); b) field-assisted ($H = 32\ kA/m$; $f = 450\ kHz$) degradation of TC by MNFs (black circles) after two applications of 60 min each. The noisy jumps in the curve at t = 2 and 4 h indicate the times when the field was switched ON/OFF, due to transient electronic interference in the spectrophotometer.

Although the application of the magnetic field could not be maintained continuously during the whole experiment of ≈ 50 h due to limitations in our experimental setup, it is clear that after the field is turned on, the degradation rate increased (noticed as a change in the slope of the degradation profile at t = 2 and 4 h) thus being an efficient way to shorten the spontaneous degradation times. The faster degradation rate observed when the applied magnetic field was turned on, particularly in the early stages of the reaction, could be due to local magnetic effects on the catalytic active centres as well as to a more obvious thermal effect of increased temperature during the experiments on the reaction kinetics. Since the MNFs have a considerable SLP = 2,295 W/g (see Figures S5 and S6 in the Supplementary Information) the local environment of the MNFs and circulating fluid increased the local temperature observed up to T = 51 °C (see Figure 8), despite the thermal insulation provided for the cooling water jacket (at constant T = 26 °C), confirming that the source of the heat are the MNFs. Even though our experimental design did not allow to identify the actual mechanism for the enhanced degradation rates, the practical implications of this effect for potential applications are clear.

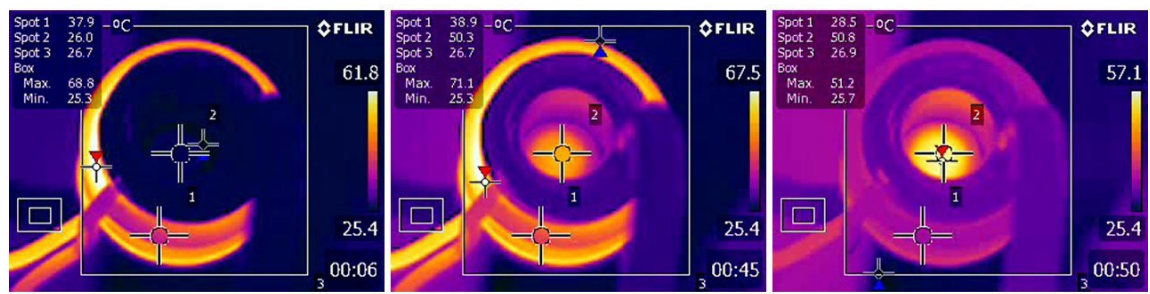

Figure 8. Thermographic screen captures from **Error! Reference source not found.** showing the heating of MNFs under an alternating magnetic field (H = 32 kA/m; $f$ = 350 kHz). The panels display the temperature distribution at different time intervals: (a) 100 s, (b) 140 s, and (c) 145 s (few seconds after field is turned off). The hotspots indicate the temperatures at the coil (spots 1 and 3) and on the sample surface (spot 2).

## 4. Discussion

The chromatographic analysis of tetracycline (TC) solutions before and after treatment with $MnFe_2O_4$-embedded MNFs revealed a complete disappearance of the TC peak (comparing to the 5 ppm TC control) following circulation through the MNF mats. As shown in Figure 9, the UHPLC-UV at 350 nm mass spectrometric monitoring of the $[M+H]^+$ ion at m/z = 445 (see also Figure S7 in the Supplementary Information), confirmed that the untreated sample (pig manure) exhibited a well-defined peak with a retention time of approximately 3.5 min. In contrast, the chromatogram corresponding to the post-treatment sample showed no detectable signal at this retention time, indicating near-complete depletion of TC from the aqueous phase. Same behaviour was observed for the final product of the pig manure after the magnetic-field driven degradation experiments, where the kinetics of degradation is accelerated.

This observation is consistent with a dual mechanistic process involving both adsorption and heterogeneous Fenton-like catalysis. Initially, TC molecules are adsorbed onto the surface of the electrospun PAN nanofibers, facilitated by hydrophobic and electrostatic interactions with the polymer and nanoparticle interfaces. Subsequently, the $MnFe_2O_4$ nanoparticles embedded within the nanofibers catalyse the decomposition of hydrogen peroxide via a surface-bound Fenton-like mechanism, generating highly reactive hydroxyl radicals (•OH). These ROS are capable of oxidatively cleaving the tetracyclic core and associated functional groups of TC, leading to its fragmentation into lower-molecular-weight species that are undetectable under the applied chromatographic conditions. The complete suppression of the TC peak in the UHPLC chromatogram thus confirms that the MNFs serve not merely as passive adsorbents but as catalytically active supports that promote the irreversible degradation of TC. These results align with the real time UV-vis detection data described in Section 3.2 and gives support to the kinetic model described below in which an initial adsorption–induction phase is followed by accelerated oxidative degradation, culminating in a saturation regime associated with catalyst deactivation and ROS depletion.

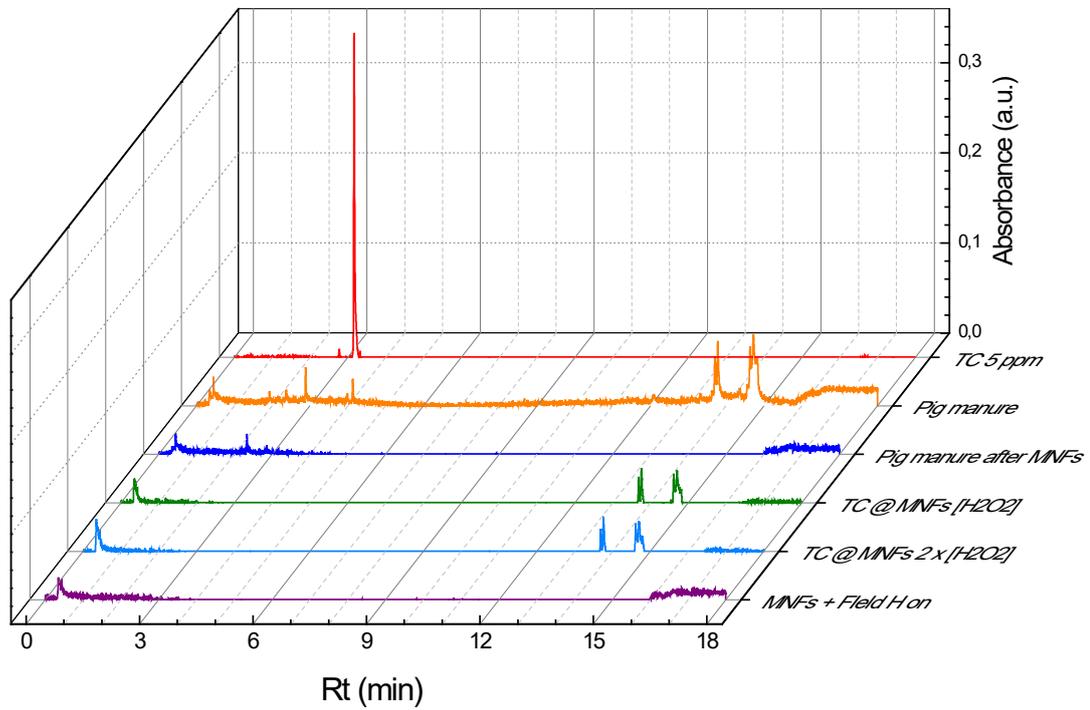

Figure 9. Chromatographic analysis of: TC control (red curve, 5 ppm TC), pig manure samples with TC before and after treatment with MnFe$_2$O$_4$-embedded MNFs with H$_2$O$_2$ addition (orange and blue lines). The disappearance of the TC peak at retention time 3.5 min, as measured by UHPLC-UV at 350 nm mass spectrometric monitoring of the [M+H]$^+$ ion at m/z = 445 can be observed in all experiments of MNFs degradation (green and light blue line) and with (violet line) applied magnetic field.

*4.2. Kinetics of TC degradation*

The Dynamic Kinetic Model (DKM) provides a time-dependent, mechanistic representation of the degradation process through differential equations, capturing the evolution of chemical species, including active catalytic sites, hydrogen peroxide, and ROS. It also accounts for the inactivation and oxidation state changes of active sites over time, reflecting dynamic reaction kinetics. The reaction equation for the kinetic modelling of TC degradation involves a series of Fenton-like reactions for the generation of reactive oxygen species, primarily $\bullet OH$:

$$Fe^{2+} + H_2O_2 \rightarrow Fe^{3+} + \bullet OH + OH^- \qquad (1)$$

$$Fe^{3+} + H_2O_2 \rightarrow Fe^{2+} + \bullet OH_2 + H^+ \qquad (2)$$

$$TC + \bullet OH \rightarrow degraded\ products \qquad (3)$$

This approach explicitly includes terms for the generation of ROS, the consumption of H$_2$O$_2$, and the inactivation of active sites due to changes in oxidation states. The model variables are a) the concentration of active sites at time t, $[A(t)]$; b) the concentration of hydrogen peroxide at time t, $[H_2O_2(t)]$; and c) the concentration of reactive oxygen species at time t, $[ROS(t)]$. The set of differential equations by that describe the evolution of tetracycline as a function of time, $[TC(t)]$, are

$$\frac{d[A(t)]}{dt} = -k_{inact}\,[A(t)] \qquad (4)$$

$$\frac{d[H_2O_2(t)]}{dt} = -k_{gen\_eff}\,[H_2O_2(t)]\,[A(t)] \tag{5}$$

$$\frac{d[ROS(t)]}{dt} = k_{gen\_eff}\,[H_2O_2(t)]\,[A(t)] - k_{deg}\,[\bullet OH(t)]\,[TC(t)] \tag{6}$$

$$\frac{d[TC(t)]}{dt} = -k_{deg}\,[\bullet OH(t)]\,[TC(t)] \tag{7}$$

where

$k_{gen}$: Rate constant for the generation of •OH from $H_2O_2$.

$k_{deg}$: Rate constant for the degradation of tetracycline by •OH.

$k_{inact}$: Rate constant for the inactivation of active sites.

These equations describe the degradation of TC by ROS (mainly •OH) generation produced by $H_2O_2$ at a decreasing number of active sites on the MNPs surface. The effective generation rate constant, $k_{gen\_eff}$, is determined based on an induction time defined as:

If $t <$ induction time $\Rightarrow k_{gen\_eff} = k_{gen\_initial}$

If $t \geq$ induction time $\Rightarrow k_{gen\_eff} = k_{gen}$

To implement this model, we numerically solved these equations through a simple Runge-Kutta based code (provided in the Supplementary Material). The induction period, introduced to account for the initial time window in which the degradation velocity is lower, can be explained by a "peeling" effect of the PAN polymer of the nanofibers that cover a fraction of the MNPs initially, hindering direct contact with the circulating fluid. As shown from HPLC data, the final carrier liquid contained small concentrations of the polymer, supporting this mechanism.

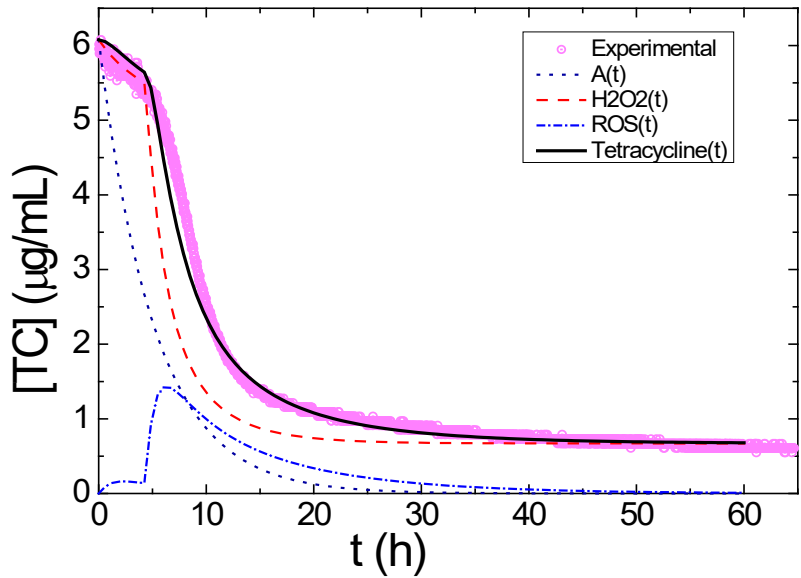

Figure 10. Experimental (pink dots) and simulated data (black line) for tetracycline (TC) degradation kinetics, using the dynamic kinetic model for Fenton-like reactions, including the generation of reactive oxygen species (ROS), active site inactivation, and hydrogen peroxide consumption (Equations 4-7). An induction period is incorporated to reflect the delayed exposure of active sites due to polymer peeling from the nanofibers.

An induction period was added to the model to account for the delayed exposure of active sites. This delay is linked to the gradual stripping of polymer layers from magnetic nanofibers, which initially hinders the full catalytic activity of the system. Figure 11 illustrates the degradation efficiency (%) as a function of time (hours) for the experiment conducted after the addition of $H_2O_2$ (2% v/v). The initial generation rate constant of reactive oxygen species (ROS), $k_{gen\_initial}$, is set to 0.02, indicating a slower rate of ROS production during the induction period, likely due to limited active site availability. After the induction period, the effective ROS generation rate $k_{gen}$ increases to 0.15, showing the catalytic activity as more active sites become exposed. The degradation rate constant, $k_{deg}$ = 0.45 is a measure of ROS mechanism breaking down tetracycline molecules. The value for $k_{inact}$ = 0.01 that yielded the best fit suggests a gradual inactivation of active sites over time, possibly due to the depletion of active sites on the surface ($Mn^{2+}$, $Fe^{2+}$) during the oxidation process. An induction time of 5.3 h suggests a delay before optimal degradation conditions are reached, supporting the hypothesis of a gradual polymer "peeling" effect exposing the embedded magnetic nanoparticles. These parameter values achieve the best fit to the experimental data.

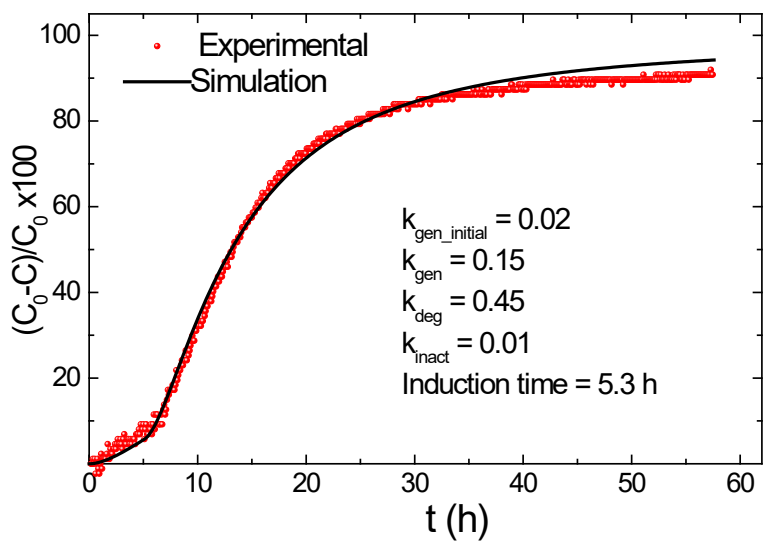

Figure 11. Experimental (red dots) and simulated data (block line) for tetracycline (TC) degradation kinetics, using the dynamic kinetic model for Fenton-like reactions, including the generation of reactive oxygen species (ROS), active site inactivation, and hydrogen peroxide consumption (Equations 4-7). An induction period is incorporated to reflect the delayed exposure of active sites due to polymer peeling from the nanofibers.

Recently, manganese, which functions similarly to iron in heterogeneous catalysis, has attracted considerable interest.[16]. $MnFe_2O_4$ is a stable spinel crystal structure with variable Fe(II, III) and Mn(II, III) species that can provide an effective $MnFe_2O_4/H_2O_2$ catalyst for different AOP applications. The magnetic nature of $MnFe_2O_4$, with relatively low magnetic anisotropy ($K_1$ = 2-8 kJ/m$^3$, ref.[17]) compared to magnetite $Fe_3O_4$, and large magnetic moment at room temperature ($M_S$ = 50-60 Am$^2$kg$^{-1}$, refs.[18]) make possible to use this material for remote magnetic heating to accelerate the reaction kinetics. However, the collection of magnetic nanoparticles when they are in colloidal suspension poses several technical challenges and increases the costs of large-scale applications.

To improve the catalytic efficiency by minimizing particle aggregation, and to avoid subsequent recollection of the magnetic catalysts, in this work we designed a Fe/Mn-based heterogeneous Fenton-like catalysts by integrating the MNPs onto a polymeric nanofiber and fabricating a maneuverable magnetic patch that can be magnetically activated by a remote alternating magnetic field. We explored the use of these magnetic nanofibers (MNFs) as a double-effect heterogeneous Fenton catalyst for the degradation of TC. Our experimental setup allowed us to investigate the MNFs for its Fenton catalytic activity for TC decomposition under alternating magnetic field in the presence of $H_2O_2$. We examined the influence of different $H_2O_2$ concentrations and investigated the mechanism and stability of the catalyst.

*Sample from Pig Manure*

After the extraction of the TC-containing residue from pig manure sample, a total volume of ≈200 mL was obtained for testing our degradation process. However, the concentration of TC detected by

HPLC in this 'as-cast' effluent was 8.2 ng/mL, much below our UV-vis detection limit (≈0.1-0.2 µg/mL). In order to reach a detectable concentration, the sample was concentrated by controlled solvent evaporation, reducing the total volume by a factor of 40, yielding to a 5 mL sample (the minimum amount for a two-run measurement). This process increased the tetracycline concentration in the sample, reaching a final value of 0.32 µg/mL, suitable for the degradation monitoring analysis by UV-Vis spectroscopy at the corresponding wavelength. The results of the tetracycline (TC) degradation by our MNFs are shown in Figure 12, where a clear decrease down to approximately 50% of the initial concentration is observed over a 20-hour circulation of the contaminated solution through the MNFs.

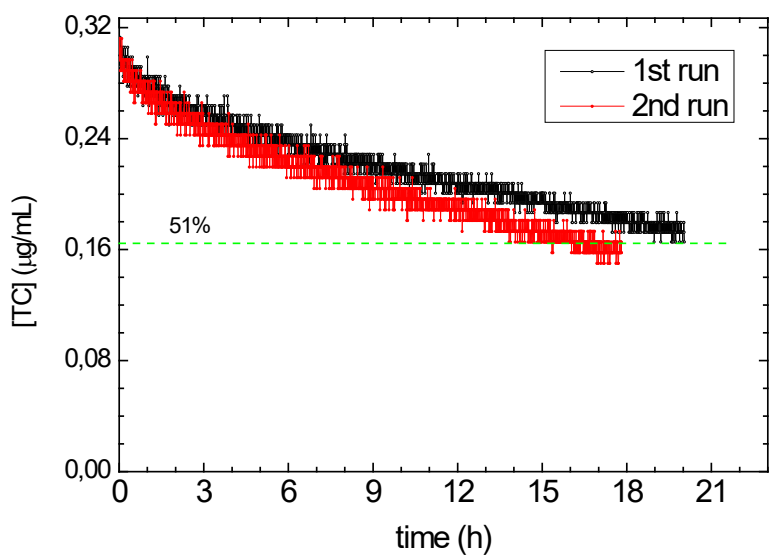

Figure 12. Degradation of tetracycline vs. time during circulation of a pig manure-derived sample through MNFs. Two independent runs (black and red curves) showed a decrease in TC concentration reaching approximately 50% of the initial value after 20 h of circulation, when the concentration (~0.16 µg/mL) reached the detection limit of the UV-Vis spectrophotometer. The sample used corresponds to a manure slurry filtrate spiked with TC and treated as an "as-cast" effluent.

Two experimental runs are presented (black and red curves), showing similar degradation profiles. The final TC concentration reached ~0.16 µg/mL, which was about the detection limit of the UV-Vis spectrophotometer used in the experiment. This explains the increased noise and data dispersion toward the end of the measurement. The slight differences between the two runs are within experimental uncertainty and are not considered significant.

This observed degradation of tetracycline by an advanced oxidation process (AOP) using our magnetic nanofibers has significant implications for applications in the treatment of livestock waste such as pig manure. The reduction of TC concentration by approximately 50% over 20 hours demonstrates the catalytic efficiency of the MNFs in generating reactive oxygen species capable of degrading persistent antibiotics. This is especially relevant given the high antibiotic loads commonly found in manure from intensive farming, which contribute to the spread of antimicrobial resistance (AMR) in the environment. The ability of the MNFs to operate under continuous flow conditions highlights their potential for integration into manure treatment systems, either in liquid effluent streams or as part of a modular purification process. Such applications could play a crucial role in reducing pharmaceutical contaminants before land application or water discharge, promoting safer agricultural practices and minimizing environmental and public health risks.

## 5. Conclusions

Through the development of a novel system composed of MnFe$_2$O$_4$@PAN magnetic nanofibers of high stability and catalytic efficiency. This study has demonstrated the effective degradation of TC in pig manure wastewater. The MNPs displayed good performance as heterogeneous Fenton-like catalysts across multiple adsorption-oxidation cycles. A dynamic kinetic model incorporating ROS generation, H$_2$O$_2$ consumption, and active site inactivation could fit the observed degradation profiles. Furthermore, the application of an alternating magnetic field could enhance the reaction rate due to localized heating effects. The robustness against extreme pH (4-14), flexibility, and easiness to handle make them interesting for potential construction of filters and coatings inside pipes. The findings highlight the potential of MNFs for scalable wastewater treatment applications, offering a reusable and efficient alternative to conventional homogeneous Fenton processes while minimizing iron sludge formation and broadening the operational pH range.

**Author Contributions:** Conceptualization, Gerardo Goya; Data curation, Berta Centro, Marco Antonio Morales Ovalle and Vanina Franco; Formal analysis, Vanina Franco and Gerardo Goya; Funding acquisition, Gerardo Goya; Investigation, Berta Centro and Marco Antonio Morales Ovalle; Methodology, Marco Antonio Morales Ovalle, Vanina Franco, Jesus Fuentes Garcia and Gerardo Goya; Supervision, Jesus Fuentes Garcia and Gerardo Goya; Writing – original draft, Berta Centro; Writing – review & editing, Marco Antonio Morales Ovalle, Vanina Franco, Jesus Fuentes Garcia and Gerardo Goya. All authors have read and agreed to the published version of the manuscript.

**Funding:** This research was funded by EU Commission through projects MSCA-RISE #101007629 (NESTOR) and by Agencia Estatal de Investigacion and EU through project (PCI2024-153438, #501100011033), M-ERA-NET2023 (ROSSCA).

**Acknowledgments:** We sincerely thank **Grupo Jorge S.L.** for providing the purine samples from a pig farm in Aragón, Spain, which were essential for this study. The authors also acknowledge the access to the facilities of the Servicio General de Apoyo a la Investigación-SAI, Universidad de Zaragoza.

**Conflicts of Interest:** The authors declare no conflicts of interest.

## Abbreviations

The following abbreviations are used in this manuscript:

| | |
|---|---|
| **AOPs** | Advanced Oxidation Processes |
| **SDGs** | Sustainable Development Goals |
| **AMF** | Alternating Magnetic Field |
| **EQS** | Environmental Quality Standards |
| **DKM** | Dynamic Kinetic Model |
| **MNPs** | Magnetic Nanoparticles |
| **MNFs** | Magnetic Nanofibers |
| **EDS** | Energy-Dispersive X-ray Spectroscopy |
| **•OH** | Hydroxyl Radical |
| **PAN** | Polyacrylonitrile |
| **PAHs** | Polycyclic Aromatic Hydrocarbons |

## References

1. Yang, J.; Li, J.; van Vliet, M. T.; Jones, E. R.; Huang, Z.; Liu, M.; Bi, J., Economic risks hidden in local water pollution and global markets: A retrospective analysis (1995–2010) and future perspectives on sustainable development goal 6. *Water Research* **2024,** *252*, 121216.

2.	Singh, N.; Poonia, T.; Siwal, S. S.; Srivastav, A. L.; Sharma, H. K.; Mittal, S. K., Challenges of water contamination in urban areas. In *Current directions in water scarcity research*, Elsevier: 2022; Vol. 6, pp 173-202.

3.	(a) Zaitseva, N. V.; Shur, P. Z.; Atiskova, N. G.; Kiryanov, D. A.; Kamaltdinov, M. R., Human health hazards associated with tetracycline drugs residues in food. *International Journal of Advanced Research* **2014,** *2* (8), 488-495; (b) Wang, H.; Wu, Y.; Feng, M.; Tu, W.; Xiao, T.; Xiong, T.; Ang, H.; Yuan, X.; Chew, J. W., Visible-light-driven removal of tetracycline antibiotics and reclamation of hydrogen energy from natural water matrices and wastewater by polymeric carbon nitride foam. *Water Research* **2018,** *144*, 215-225.

4.	Xu, L.; Zhang, H.; Xiong, P.; Zhu, Q.; Liao, C.; Jiang, G., Occurrence, fate, and risk assessment of typical tetracycline antibiotics in the aquatic environment: A review. *Science of the total Environment* **2021,** *753*, 141975.

5.	Babu Ponnusami, A.; Sinha, S.; Ashokan, H.; V Paul, M.; Hariharan, S. P.; Arun, J.; Gopinath, K. P.; Hoang Le, Q.; Pugazhendhi, A., Advanced oxidation process (AOP) combined biological process for wastewater treatment: A review on advancements, feasibility and practicability of combined techniques. *Environmental Research* **2023,** *237*, 116944.

6.	Wang, J. L.; Xu, L. J., Advanced oxidation processes for wastewater treatment: formation of hydroxyl radical and application. *Critical reviews in environmental science and technology* **2012,** *42* (3), 251-325.

7.	Luo, H.; Zeng, Y.; He, D.; Pan, X., Application of iron-based materials in heterogeneous advanced oxidation processes for wastewater treatment: A review. *Chemical Engineering Journal* **2021,** *407*, 127191.

8.	Cheng, M.; Zeng, G.; Huang, D.; Lai, C.; Liu, Y.; Xu, P.; Zhang, C.; Wan, J.; Hu, L.; Xiong, W.; Zhou, C., Salicylic acid–methanol modified steel converter slag as heterogeneous Fenton-like catalyst for enhanced degradation of alachlor. *Chemical Engineering Journal* **2017,** *327*, 686-693.

9.	Ramachanderan, R.; Schaefer, B., Tetracycline antibiotics. *ChemTexts* **2021,** *7* (3), 18.

10.	Phouthavong, V.; Yan, R.; Nijpanich, S.; Hagio, T.; Ichino, R.; Kong, L.; Li, L., Magnetic adsorbents for wastewater treatment: advancements in their synthesis methods. *Materials* **2022,** *15* (3), 1053.

11.	Moreno Maldonado, A. C.; Winkler, E. L.; Raineri, M.; Toro Córdova, A.; Rodríguez, L. M.; Troiani, H. E.; Mojica Pisciotti, M. L.; Mansilla, M. V.; Tobia, D.; Nadal, M. S., Free-radical formation by the peroxidase-like catalytic activity of $MFe_2O_4$ (M= Fe, Ni, and Mn) nanoparticles. *The Journal of Physical Chemistry C* **2019,** *123* (33), 20617-20627.

12.	Fuentes-García, J. A.; Sanz, B.; Mallada, R.; Ibarra, M. R.; Goya, G. F., Magnetic nanofibers for remotely triggered catalytic activity applied to the degradation of organic pollutants. *Materials & Design* **2023,** *226*, 111615.

13.	Ziegler, R.; Ilyas, S.; Mathur, S.; Goya, G. F.; Fuentes-García, J. A., Remote-controlled activation of the release through drug-loaded magnetic electrospun fibers. *Fibers* **2024,** *12* (6), 48.

14.     Goya, G. F.; Mayoral, A.; Winkler, E.; Zysler, R. D.; Bagnato, C.; Raineri, M.; Fuentes-García, J. A.; Lima, E., Next generation of nanozymes: A perspective of the challenges to match biological performance. *Journal of Applied Physics* **2021,** *130* (19).

15.     Fuentes-García, J. s. A.; Carvalho Alavarse, A.; Moreno Maldonado, A. C.; Toro-Córdova, A.; Ibarra, M. R.; Goya, G. F. n., Simple sonochemical method to optimize the heating efficiency of magnetic nanoparticles for magnetic fluid hyperthermia. *ACS omega* **2020,** *5* (41), 26357-26364.

16.     (a) Li, Y.; Qu, J.; Gao, F.; Lv, S.; Shi, L.; He, C.; Sun, J., In situ fabrication of Mn3O4 decorated graphene oxide as a synergistic catalyst for degradation of methylene blue. *Applied Catalysis B: Environmental* **2015,** *162*, 268-274; (b) Duan, L.; Wang, Z.; Hou, Y.; Wang, Z.; Gao, G.; Chen, W.; Alvarez, P. J. J., The oxidation capacity of Mn3O4 nanoparticles is significantly enhanced by anchoring them onto reduced graphene oxide to facilitate regeneration of surface-associated Mn(III). *Water Research* **2016,** *103*, 101-108; (c) Wan, Z.; Wang, J., Degradation of sulfamethazine using Fe3O4-Mn3O4/reduced graphene oxide hybrid as Fenton-like catalyst. *Journal of Hazardous Materials* **2017,** *324*, 653-664.

17.     Sanz, B.; Cabreira-Gomes, R.; Torres, T. E.; Valdés, D. P.; Lima, E., Jr.; De Biasi, E.; Zysler, R. D.; Ibarra, M. R.; Goya, G. F., Low-Dimensional Assemblies of Magnetic MnFe2O4 Nanoparticles and Direct In Vitro Measurements of Enhanced Heating Driven by Dipolar Interactions: Implications for Magnetic Hyperthermia. *ACS Applied Nano Materials* **2020,** *3* (9), 8719-8731.

18.     (a) Ma, Y.; Xu, X.; Lu, L.; Meng, K.; Wu, Y.; Chen, J.; Miao, J.; Jiang, Y., Facile synthesis of ultrasmall MnFe2O4 nanoparticles with high saturation magnetization for magnetic resonance imaging. *Ceramics International* **2021,** *47* (24), 34005-34011; (b) Li, J.; Yuan, H.; Li, G.; Liu, Y.; Leng, J., Cation distribution dependence of magnetic properties of sol–gel prepared MnFe2O4 spinel ferrite nanoparticles. *Journal of Magnetism and Magnetic Materials* **2010,** *322* (21), 3396-3400.